\documentclass[
 aip,
rsi,
 amsmath,amssymb,
 reprint,%
]{revtex4-1}

\usepackage{graphicx}
\usepackage{dcolumn}
\usepackage{bm}

\usepackage[utf8]{inputenc}
\usepackage[T1]{fontenc}
\usepackage{mathptmx}
\usepackage{etoolbox}

\makeatletter
\def\@email#1#2{%
 \endgroup
 \patchcmd{\titleblock@produce}
  {\frontmatter@RRAPformat}
  {\frontmatter@RRAPformat{\produce@RRAP{*#1\href{mailto:#2}{#2}}}\frontmatter@RRAPformat}
  {}{}
}%
\makeatother
\begin{document}

\preprint{AIP/123-QED}

\title[Longitudinal tapering in gas jets for increased efficiency of 10-GeV class laser plasma accelerators]{Longitudinal tapering in gas jets for increased efficiency of 10-GeV class laser plasma accelerators}
\author{R. Li$^*$}
\affiliation{Lawrence Berkeley National Laboratory, 1 Cyclotron Road, Berkeley CA 94720, USA}
\affiliation{Department of Nuclear Engineering, University of California, Berkeley, California 94720, USA}
\email{raymondli@lbl.gov}
\author{A. Picksley}
\affiliation{Lawrence Berkeley National Laboratory, 1 Cyclotron Road, Berkeley CA 94720, USA}
\author{C. Benedetti}
\affiliation{Lawrence Berkeley National Laboratory, 1 Cyclotron Road, Berkeley CA 94720, USA}
\author{F. Filippi}
\affiliation{Lawrence Berkeley National Laboratory, 1 Cyclotron Road, Berkeley CA 94720, USA}
\affiliation{ENEA Nuclear Department-C. R. Frascati, Via Enrico Fermi 45, 00044 Frascati, Italy}
\author{J. Stackhouse}
\affiliation{Lawrence Berkeley National Laboratory, 1 Cyclotron Road, Berkeley CA 94720, USA}
\affiliation{Department of Nuclear Engineering, University of California, Berkeley, California 94720, USA}
\author{L. Fan-Chiang}
\affiliation{Lawrence Berkeley National Laboratory, 1 Cyclotron Road, Berkeley CA 94720, USA}
\affiliation{Applied Science and Technology, College of Engineering, University of California, Berkeley, California 94720, USA}
\author{H. E. Tsai}
\author{K. Nakamura}
\affiliation{Lawrence Berkeley National Laboratory, 1 Cyclotron Road, Berkeley CA 94720, USA}
\author{C. B. Schroeder}
\affiliation{Lawrence Berkeley National Laboratory, 1 Cyclotron Road, Berkeley CA 94720, USA}
\affiliation{Department of Nuclear Engineering, University of California, Berkeley, California 94720, USA}
\affiliation{Applied Science and Technology, College of Engineering, University of California, Berkeley, California 94720, USA}
\author{J. van Tilborg}
\author{E. Esarey}
\author{C. G. R. Geddes}
\author{A. J. Gonsalves}
\affiliation{Lawrence Berkeley National Laboratory, 1 Cyclotron Road, Berkeley CA 94720, USA}

\date{\today}

\begin{abstract}
Modern laser plasma accelerators (LPAs) often require plasma waveguides tens of cm long to propagate a high-intensity drive laser pulse. Tapering the longitudinal gas density profile in 10~cm scale gas jets could allow for single stage laser plasma acceleration well beyond 10~GeV with current petawatt-class laser systems. Via simulation and interferometry measurements, we show density control by longitudinally adjusting the throat width and jet angle. Density profiles appropriate for tapering were calculated analytically and via particle-in-cell (PIC) simulations, and were matched experimentally. These simulations show that tapering can increase electron beam energy using 19~J laser energy from $\sim$9~GeV to $>$12~GeV in a 30~cm plasma, and the accelerated charge by an order of magnitude.

This paper was published in Review of Scientific Instruments on April 11, 2025 DOI: https://doi.org/10.1063/5.0250698
\end{abstract}

\maketitle

\section{Introduction}
\label{sec:intro}
Laser plasma accelerators (LPAs) \cite{Tajima1979,esarey2009} are a promising technology for building compact accelerators because they can generate electric fields 10-100~GeV/m \cite{esarey2009}, which is orders of magnitude higher than conventional accelerators. In an LPA, a high-intensity laser beam's ponderomotive force excites plasma waves, which generate large electric fields inside a plasma. These electric fields are used to accelerate electrons to high energies. LPAs are studied for many applications, including free-electron lasers \cite{wang2021free}, high-energy particle colliders \cite{schroeder2010physics}, and nuclear detection \cite{geddes2015compact}. Maximizing efficiency, defined as the fraction of laser energy transferred to the electron beam, is crucial for these applications.

Generating 10-GeV-class electron bunches requires an accelerator tens of cm long with currently available petawatt (PW) laser systems \cite{esarey2009}. Since the Raleigh range for a typical PW-class drive laser is $\sim$1~cm, diffraction limits the range where the drive laser is intense enough to drive an LPA. Different techniques have been developed to prevent diffraction, including relativistic self-guiding \cite{borisov1992observation, krushelnick1997plasma} and pre-formed plasma channel waveguides \cite{durfee1993light, zigler1996optical}, which can be created using methods such as hydrodynamic expansion of collisonally heated plasmas \cite{durfee1995development, volfbeyn1999guiding, ditmire1998plasma, kumarappan2005guiding}, ponderomotively formed channels \cite{ting1997plasma, krushelnick1997plasma}, capillary discharges \cite{butler2002guiding, spence2003gas, leemans2006gev, leemans2014multi, Gonsalves2019}, or hydrodynamic optical field ionization (HOFI) \cite{shalloo2018hydrodynamic, shalloo2019low, picksley2020meter, feder2020self}.
Gas jets are a common plasma source for a large variety of laser-plasma interaction experiments, including LPAs in the self-guided or guided regimes. Gas jets based on elongated converging-diverging (de Laval) nozzles have been demonstrated at lengths up to 30~cm \cite{miao2022, shrock2022meter, miao2024meter}. We have recently employed a 30-cm long gas jet to generate quasimonoenergetic electron beams at 9.2~GeV, with charge extending beyond 10~GeV using HOFI channels and a 21 J, 40 fs drive laser \cite{Picksley2024}.

In a guided LPA, the energy achievable in a single LPA stage is often limited by dephasing. This occurs when the accelerating electron bunch has a higher velocity than the drive laser, which travels at the group velocity of light in a plasma $v_g < c$. Eventually, the electron bunch advances in phase and moves into a decelerating region of the plasma wakefield. Given fixed laser parameters and acceleration length, dephasing can be mitigated by introducing a density upramp along the accelerating region \cite{Suk2003, pukhov2008control, Rittershofer2010}. An increasing plasma density decreases the local plasma wavelength. As the electrons travel through the LPA and move closer to the laser pulse, a decreasing plasma wavelength can keep the electron bunch in the accelerating portion of the wake. Density gradients have been shown to increase electron energy in mm-scale gas targets \cite{Guillaume2015, Aniculaesei2019} at sub-GeV energies. To apply such gradients for LPAs at 10~GeV and beyond would require targets that are tens of cm long, with density tapering not previously achieved.

In this paper, we experimentally demonstrate that the tapering profiles derived analytically and via simulations can be achieved through simple adjustments to the 30~cm gas jet that was previously used (not tapered) to achieve electron energies up to 10~GeV \cite{Picksley2024}. To support our experimental gas density profile measurements, we simulated gas jet nozzles in OpenFOAM \cite{jasak2007openfoam} and show good agreement with the experimental measurements. Particle-in-cell (PIC) simulations, performed with the code INF\&RNO \cite{benedetti2012quasi, benedetti2017accurate}, indicate that using 19~J of laser energy\cite{Picksley2024} , linear tapering of a 30~cm-long gas jet could result in a 9$\times$ increase in charge and a 32\% increase in electron energy compared to the results obtained in the untapered case.

\begin{figure*}[t]
\centering
\includegraphics[width=\linewidth]{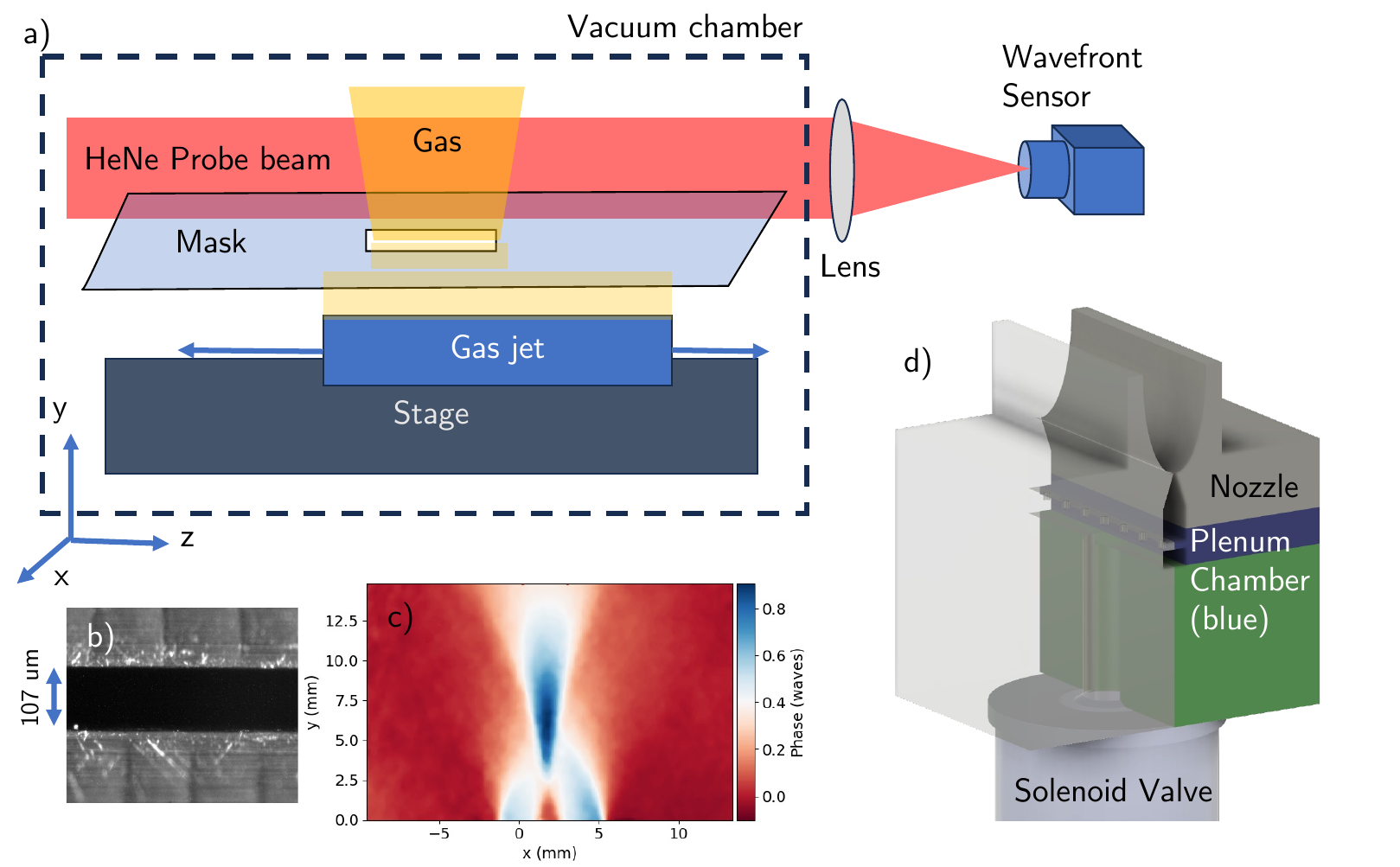}
\caption{a) Gas jet density measurement setup. The phase offset of the probe beam ($\sim$3~cm diameter) after it passes through the gas plume was measured by a wave front sensor. b) Example microscope image of gas jet throat used for width measurements. c) Raw phase data measured from a representative shot. A 3D model of a small section of the 30~cm elliptical gas jet is shown in d).}\label{fig:experimentalSetup}
\end{figure*}

\section{\label{sec:experiment}Gas Density Measurements}

\begin{figure}[t]
\centering
\includegraphics[width=\linewidth]{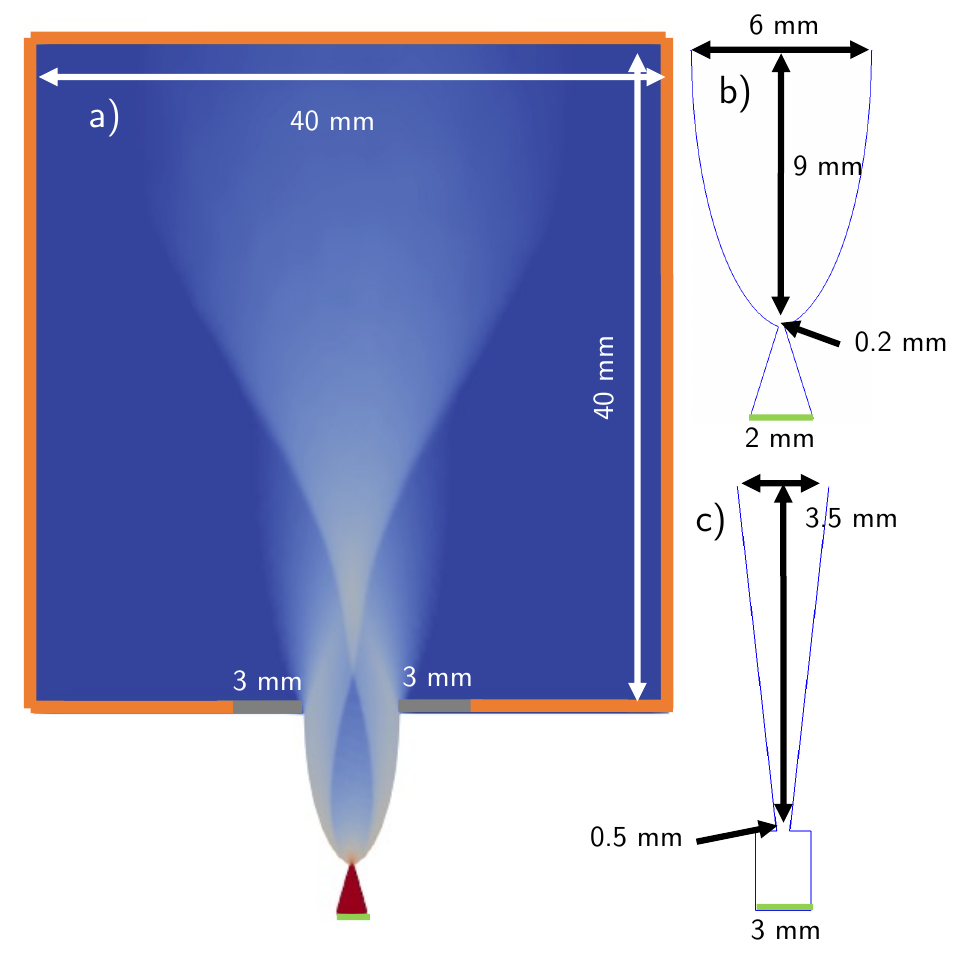}
\caption{The simulation domain is shown in a). Inlet boundary conditions are marked in green, outlets are marked in orange, and additional walls near the exit of the nozzle are marked in gray. We simulated both an elliptical and straight nozzle shape, shown in b) and c) respectively.}\label{fig:simDomain}
\end{figure}

In our experiments, three different gas jets similar in design to Ref.~\onlinecite{krishnan2012linear, krishnan2012novel, mittelberger2019laser, Ocean2021,miao2022} were used: 
(i) a 30~cm long gas jet with 10 equally spaced solenoid valves \cite{Picksley2024} that fed into an elongated de Laval nozzle with an elliptical cross section \cite{Ocean2021} (shown in Figure \ref{fig:simDomain}b); (ii) a 5~cm long jet that is otherwise identical except that it uses 4 valves; and (iii) a 2~cm jet with 1 solenoid valve and a straight diverging section shown in Figure \ref{fig:simDomain}c. 

The 30~cm gas jet was constructed (See Figure \ref{fig:experimentalSetup}d for a 3D model) by connecting 10 solenoid valves to a plenum chamber that extends along the entire jet length, which dampened any potential shock waves and evenly distributed the gas along the gas jet length. This plenum chamber was then connected to the nozzle of the gas jet, which is an elongated converging-diverging nozzle with cross section shown in Figure \ref{fig:simDomain}b.

The converging section of the elliptical nozzles have a 2~mm inlet, which linearly converges to a 200 $\mu$m throat. Its diverging section consists of elliptical arc walls with a major axis along the center of the nozzle and minor axis at the exit of the nozzle. The straight nozzle has a rectangular inlet section with a length of 4~mm and a width of 3~mm. The diverging section starts from a throat 0.5~mm wide and expands linearly to a width of 3.5~mm. 

Our gas density measurement setup is shown in Figure \ref{fig:experimentalSetup}a. The gas jet is mounted on a stage that moves longitudinally along the optical axis. An aluminum plate blocked all but a short section of the gas jet, which allowed the longitudinal component of the gas density profile to be measured. The plate contained plastic inserts that diverted gas flow from the masked region of the gas jet towards the ends of the jet to not disrupt the unmasked flow. To verify that the plate did not significantly impact gas flow, we compared the measured gas density of the entire 30~cm gas jet without the plate to the gas jet with the plate. We found no significant differences in the shape of the gas density profile or in the quantitative gas density. A 633~nm HeNe continuous wave (CW) laser was used as a probe beam. A high-resolution wavefront sensor (\textit{Imagine Optic HASO LIFT 680}) was used to measure the phase offset caused by the different index of refraction between gas and vacuum.

To measure the longitudinal density profile of the gas jet, the phase shifts of the probe beam were measured with the gas jet at different positions on the stage. The backing pressure was set to $1.48 \times 10^6$~Pa. At this pressure, we measured that the valves began to open at $\sim$3.5~ms after the initial trigger and reached steady state $\sim$1~ms later. We took density data at 4.5~ms after the trigger to ensure that timing fluctuations did not affect the gas density.

We found that ray deflection due to the gradient of index of refraction in the transverse directions was not a significant source of error. The misalignment of the gas jet with respect to the laser, which we estimate to be less than 3.3~mrad, was also an insignificant source of error.

The gas density can be calculated with the formula \cite{couperus2016}
\begin{equation}
n(x,y) = \frac{2 \epsilon_0}{\alpha} \frac{\phi(x,y) \lambda}{2\pi L},
\end{equation}
where $n$ is the number density, $\phi$ is the phase offset in radians, $\lambda$ is the probe beam wavelength, $L$ is the length of the hole in the mask, $\alpha$ is the polarizability of the gas, and $\epsilon_0$ is the permittivity of free space. We used $N_2$, which has an $\alpha$ of $19.62 \times 10^{-41}$ $\mathrm{F} \cdot \mathrm{m}^2$ at 633~nm \cite{couperus2016}.

\section{\label{sec:simulation}Simulation Methods}

Gas jet nozzles were simulated in OpenFOAM \cite{jasak2007openfoam}. Since the exact shape of the nozzle has a large impact on the transverse gas density profile \cite{Ocean2021}, computational fluid dynamics simulations can allow us to understand the effects of geometry changes on gas density distribution. Nozzles were simulated in 2D due to the lengthwise symmetry of the system. The valves are not simulated because 3D simulations with a complex geometry would be required, greatly increasing the computation cost. The simulation domain with our gas jet geometries is shown in Figure \ref{fig:experimentalSetup}e. 

The simulations used hydrogen gas with a Sutherland viscosity model, given by 
\begin{equation}
\mu = \frac{A_s T^{1.5}}{T + T_s},
\end{equation}
where $\mu$ is the dynamic viscosity in Pa $\cdot$ s, $T$ is temperature in K, and $A_s$ and $T_s$ are constants. We used $A_s = 8.59 \times 10^{-7} Pa \cdot s ~ K^{-0.5}$ and $T_s = 273 K$ in our simulations.
The standard k-$\epsilon$ RAS turbulence model \cite{launder1974} with constants from Ref. \onlinecite{el1983k} was used to model turbulence.

Meshes for the simulations were created using Gmsh \cite{geuzaine2008gmsh}. A quadrilateral mesh was made using the Frontal-Delauany method \cite{rebay1993}. The mesh sizes are approximately $r_t/3$ near the throat, $r_t/1.5$ close to the walls of the nozzle and the inlet, 200~$\mathrm{\mu}$m in the center of the nozzle and directly to the right of the jet, and 1~mm near the outlet, with an approximately linear mesh size gradient at all other points in the simulation domain. $r_t$ refers to the throat width of the nozzle. 

The inlet boundary condition was set to a constant uniform pressure, with a value chosen to quantitatively match experimental data. This was necessary because the pressure at the inlet of the nozzle was not known in experiment. Instead, the experimental backing pressure was set at the valves of the gas jet. In the simulations that most closely matched experimental data, the simulated elliptical nozzle had an inlet pressure of 220~kPa and the simulated straight nozzle had an inlet pressure of 96.5~kPa. 

To validate the robustness of our simulation results against changes in the inlet pressure, we measured gas density profiles at 2 different experimental valve pressures, 0.24~MPa and 1.48~MPa with the elliptical nozzle shape. We ran simulations that quantitatively matched the measured gas densities by adjusting the inlet boundary condition pressures to 35.5~kPa and 220~kPa, respectively. The ratio of simulated inlet pressures is the same as experimental valve pressures. The density peak moves from $\sim$4~mm above the nozzle at an experimental valve pressure of 0.24~MPa to $\sim$6~mm at 1.48~MPa, as shown in Figure \ref{fig:exp-vs-sim}a. We observe good agreement in the density profile at both pressures and in the shift of the density peak, validating our simulated results.

The outlet boundary condition was an outflow boundary condition with pressure at infinity set to 1 Pa. The initial vacuum pressure was set to 100 Pa for numerical stability. We found no significant differences between simulations with different initial vacuum pressures or outlet pressures between 1-1000 Pa. 

The solver RhoCentralFoam was used. For each simulation, the mass density was extracted once the simulation reached approximately steady state, which occurs at $\sim$0.1~ms. The number density profile was then calculated and compared with experimental results.

\section{Results and Discussion}

\begin{figure}[t!]
\centering
\includegraphics[width=\linewidth]{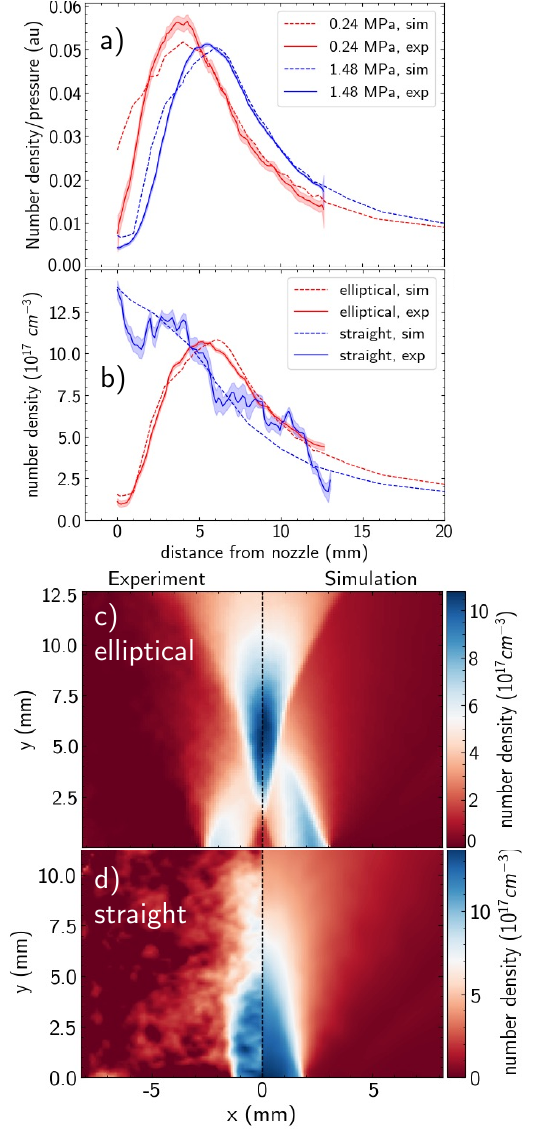}
\caption{Density lineouts from simulations and experimental data of the elliptical nozzle shape for two different pressures are shown in a). The experimental and simulated gas density profiles at 1.48~MPa are shown side by side in c) and d), for the elliptical and straight nozzle shape, respectively. Since shot to shot fluctuations in the laser were the dominant source of noise, the lower length of the 2~cm jet results in it having increased noise compared to the 5~cm jet. A lineout at x = 0 is shown for both shapes in b), with experimental measurements in solid lines and simulation results in dashed lines.}\label{fig:exp-vs-sim}
\end{figure}

\begin{figure*}[t!]
\centering
\includegraphics[width=\textwidth]{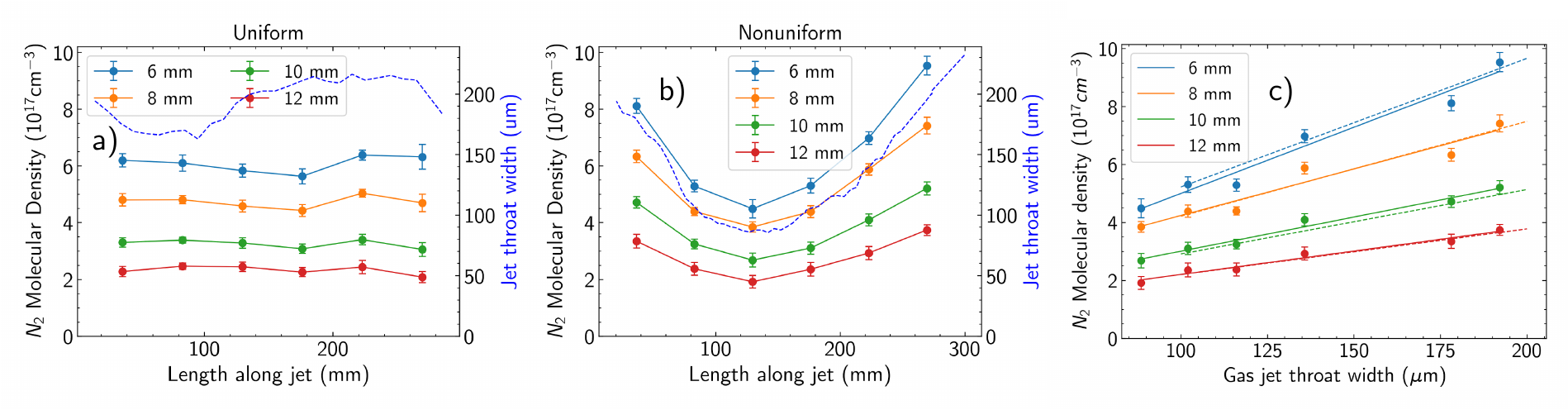}
\caption{Measured longitudinal density profile of the uniform throat width and nonuniform throat width configuration are shown in a) and b), respectively. Dashed lines plot nozzle throat width. The linear correlation between nozzle throat width and molecular density is shown in c) for different heights above the nozzle exit. Experimental regressions are shown in solid lines and simulated predictions are shown in dashed lines.}\label{fig:long-density-profiles}
\end{figure*}

When using 10~cm scale gas jets for HOFI experiments, the drive beam must be a certain height, usually $\gtrsim$5mm, away from the gas jet nozzle exit so that the channel-forming Bessel beam is not blocked by the gas jet. The minimum distance between the drive laser and the jet is lower bounded by the fact that axicon rays will be blocked if their angle is too shallow. For a 30~cm jet, the minimum angle required is $\sim$$1^\circ$, which corresponds to $\sim$5~mm over the 30~cm jet length. One goal for LPA experiments is to achieve the appropriate gas density with the lowest amount of gas possible because the pump rate of vacuum pumps limits the repetition rate in LPA experiments. Our work has been limited to 0.2~Hz for this reason. We use the simulated gas density to mass flow rate ratio as a figure of merit for repetition rate; the higher this ratio is, the higher the gas jet repetition rate can be. 

The shape of the nozzle has a large impact on the transverse gas density to mass flow rate profile \cite{Ocean2021}. We measured the gas density profile of an elliptical and straight nozzle in experiment and simulation. The density profiles of the 5~cm elliptical and the 2~cm straight gas jet are shown side by side with their respective simulations in Figure \ref{fig:exp-vs-sim}c and d. Both density profiles were measured without a mask. The 5~cm elliptical jet was used instead of the 30~cm elliptical jet because when a 30~cm jet is used with a mask, the data $\le$4~mm from the nozzle is distorted due to diffraction from the probe beam on the mask. The 5~cm elliptical jet gives the same measured density distribution as the 30~cm elliptical jet up to a constant factor that is due to differences in the mass flow rate of the valves. The simulated and experimental density profiles of the 2~cm straight jet were normalized such that the simulated mass flow rates through the elliptical and straight nozzles were identical. We observe good agreement between experiment and simulation for both the elliptical and straight nozzles.

The shape of the gas density profile for the elliptical nozzle is different than for a straight de Laval nozzle. In a straight de Laval nozzle, the density decreases monotonically with distance from the nozzle since the gas flow diverges. However, in the elliptical nozzle, the gas is focused $\sim $6~mm away from the nozzle and then starts to diverge. This caused the molecular density to peak at 6~mm and decrease approximately inversely proportional to the distance from the nozzle exit for heights $>6$~mm. The elliptical nozzle shapes gas flow such that the peak density position is optimal for LPA experiments, which enables a 30-40\% higher molecular density to mass flow rate ratio 6-12~mm away from the nozzle compared to a straight jet. This could allow for up to 30-40\% higher repetition rate.

Tapering can be achieved by exploiting the density gradient with respect to height. By tilting the gas jet such that the back of the jet is closer to the drive beam, but still $>$6~mm away, tapering can be achieved in the elliptical jet. The elliptical nozzle also reduces the risk of laser damage since the drive laser will be at least 6~mm below the nozzle.

We controlled the gas jet's longitudinal density profile via throat width adjustments. We tested two different throat width configurations with the elliptical nozzle: a uniform and a nonuniform case. The longitudinal gas density profile at different heights and the throat width profile for both configurations are shown in Figure \ref{fig:long-density-profiles}. We find that for the uniform case, the RMS longitudinal density variations are 4-6\% between 6 and 12~mm above the nozzle, while the RMS throat width variation is 9.3\%. The throat width variations in the uniform case are mostly caused by manufacturing limitations. The shot-to-shot density fluctuations can be $<$1\%, the measurement being limited by shot-to-shot fluctuations in the probe laser.
There is a linear correlation between throat width and molecular density for heights $>$6~mm above the nozzle exit, which corresponds to the region above the density maximum. This correlation was verified by simulations of nozzles with different throat widths with the same inlet pressure. In the uniform case, the reduced correlation between throat width and gas density is due to other nonuniformities, such as differences between the solenoid valves, becoming significant. 

The data indicate there are two simple methods for varying the density profile, (i) changing the throat width and (ii) tilting the gas jet. By tilting a constant throat width nozzle, we can achieve a linearly tapered density profile. For more complex density profiles, a combination of tilt and throat width adjustments can be used to generate an arbitrary longitudinal density profile.

\begin{figure}[t]
\centering
\includegraphics[width=\linewidth]{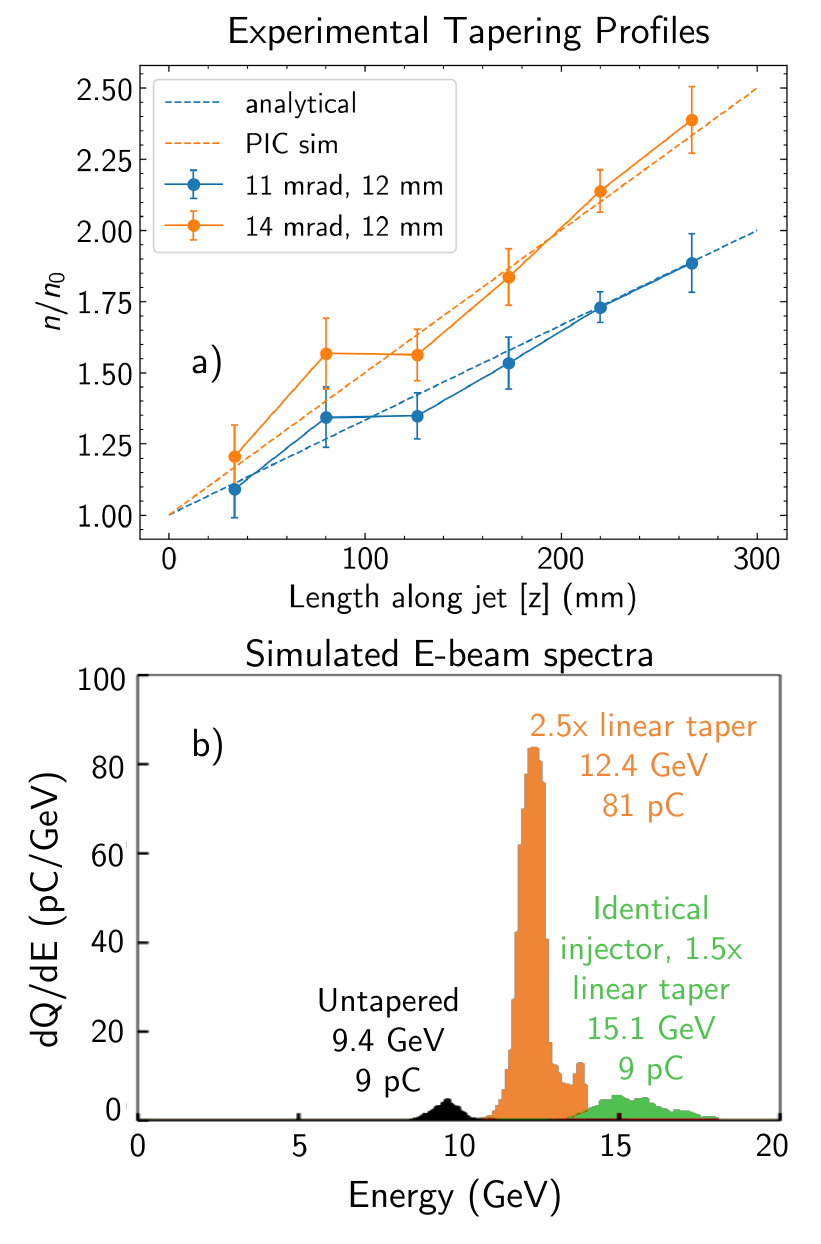}
\caption{a) The measured, normalized gas densities of the longitudinally uniform gas jet at different heights and different gas jet tilts are shown in circles with solid lines, highlighting linear tapering over 30~cm. The analytically calculated and simulated 2.5$\times$ linear tapering profiles are shown in dashed lines. b) Simulated e-beam profiles from the untapered case (black), the constant density during the injector region, followed by a 1.5$\times$ linear taper case (green), and the 2.5$\times$ linear taper case (orange) are shown.} \label{fig:tapering}
\end{figure}

The characteristic longitudinal plasma density gradient required to mitigate dephasing can be estimated with an analytical 1-D model (see, e.g., Ref.~\cite{Rittershofer2010}). We have,
\begin{align}
    \frac{n(z)}{n_0} &\approx 1 + \frac{\pi}{|\psi_0|}\frac{z}{L_d}, \quad L_d = \pi \frac{k_0^2}{k_{p0}^3},
\end{align}
where $n(z)$ is the plasma density, $n_0=n(z=0)$ is the plasma density at the upstream end of the jet, $|\psi_0|$ is the phase of the electrons behind the laser driver, $L_d$ is the dephasing length, $k_0$ is the laser wavenumber, and $k_{p0}$ is the plasma wavenumber at the upstream end of the jet. For an electron bunch loaded in the first plasma period $|\Psi_0|\sim 2\pi$, and so the expected plasma density increase beyond a dephasing length is approximately $n(z=2L_d)/n_0 \sim 2$. Density profiles corresponding to this gradient can be experimentally realized by tilting the uniform configuration jet by 11~mrad, and positioning the upstream end of the jet 12~mm below the drive laser as shown in Figure \ref{fig:tapering}a. Note that the density appears flat between $\sim$8 and $\sim$13~cm due to fluctuations in longitudinal gas density. For both the analytical and simulated tapering profile, the densities are normalized such that at $z=0$, $n/n_0 = 1$ because the molecular density can be arbitrarily scaled by changing the backing pressure or the number of valves on the jet.

We numerically investigated, by means of PIC simulations performed with the code INF\&RNO \cite{benedetti2012quasi, benedetti2017accurate}, the role of tapering in the performance of a 30~cm-long, 10~GeV-class LPA with the laser and plasma parameters discussed in Ref.~\onlinecite{Picksley2024}. 
In all cases the LPA is driven by a laser pulse with an energy of 19 J, a pulse duration of 40 fs (FWHM of the intensity), and a 53 $\mu$m laser spot, defined as the diameter where the laser intensity drops to $1/e^2$ of its maximum. The laser mode and the initial transverse density profile are the ones described in Figure 1(c) of Ref.~\onlinecite{Picksley2024}. 
The channel depth is kept constant along the channel length in these simulations. We expect this to be a reasonable approximation as, in the HOFI channel generation, the radial position of the hydrodynamic shock front is approximately independent of density.  This can be understood from Sedov-Taylor theory \cite{shalloo2018hydrodynamic}, since the shock front radius is $r_s \propto (\mathcal{E}/n_0)^{1/4}$, where $\mathcal{E}$ is the energy deposited per unit length, and  $\mathcal{E} \propto n_0$ for energy deposition using optical field ionization.
The value of the on-axis electron density at the entrance of the plasma was $1.05\times 10^{17}$~cm$^{-3}$ for all the simulations. Electron beams were generated by introducing a nitrogen dopant (1\% nitrogen / 99\% hydrogen) in the gas jet, enabling ionization-induced injection. 

We simulated an untapered case with dopant restricted to the first 12~cm of the plasma as a baseline. In this case, an electron beam with a charge of 9 pC and an average energy of 9.4~GeV, with rms energy spread of $\sim$6\% was produced. Then, a tapered case was simulated with the same dopant region as the untapered case at constant density, followed by 18~cm of a 1.5$\times$ linear taper. We chose an identical dopant region to the untapered case to avoid any effects a changing plasma density could have on laser evolution and injection, which allowed us to isolate the effects of tapering on the electron beam. In this case, the central energy rises to $>$15~GeV (Figure \ref{fig:tapering}b) with the same charge (9 pC) as the untapered case. 

We also simulated tapering profiles with a linearly increasing density along the entire gas length, which can be experimentally realized by tilting a uniform throat width gas jet. The longitudinally uniform jet tilted at 14~mrad and positioned such that the upstream end of the jet is 12~mm below the drive laser is able to create a 2.5$\times$ linear taper, as shown in Figure \ref{fig:tapering}a. In a 2.5$\times$ linear tapering simulation with the same dopant region as in the uniform case (0-12~cm), an electron beam with a charge of 225 pC, mean energy of 11~GeV, and rms energy spread of $\sim$10\% is produced. To reduce the energy spread, we simulated restricting the dopant region from $8 < z < 10$~cm. In this case, the charge dropped to 81 pC, and the energy increased to 12.4~GeV with an rms energy spread of $\sim$4\%, as shown in Figure \ref{fig:tapering}b. The increased charge arises because the increased plasma density before the injector region leads to more self-steepening, causing the laser intensity when the electrons are injected to be greater than the untapered case.

\section{Conclusion}

Longitudinal density profiles appropriate for tapering were created using a 30~cm long gas jet. We showed that by adjusting two simple parameters, the angle of the jet and the throat width, arbitrary slowly-varying density distributions could be generated. An experimental platform was constructed to measure the 3-D density distribution of 10~cm scale gas jets. An elliptical nozzle shape was experimentally measured and simulated, and was shown to have a higher peak density to mass flow rate ratio than a straight de Laval nozzle. Tapered density profiles relevant to the BELLA laser system were simulated and experimentally matched using a tilted 30~cm gas jet, proving through simulations that $>$10-GeV scale tapering is in reach for future experiments with laser drivers as low as 19~J.

\begin{acknowledgements}
This work was supported by the Defense Advanced Research Projects Agency, the Director, Office of Science, Office of High Energy Physics, of the U.S. Department of Energy under Contract No. DE-AC0205CH11231, and used the computational facilities at the National Energy Research Scientific Computing Center (NERSC). The authors thank Teo Maldonado Mancuso, Zachary Eisentraut, Mark Kirkpatrick, Federico Mazzini, Nathan Ybarrolaza, Derrick McGrew, and Chetanya Jain for technical support.\end{acknowledgements}

\section*{Author Declarations}

\subsection*{Conflicts of Interest}
The authors have no conflicts to disclose.

\subsection*{Author Contributions}

\textbf{R. Li}: Conceptualization (equal), Data curation (lead), Formal analysis (lead), Methodology (lead), Investigation (lead), Visualization (lead), Writing - original draft (lead).
\textbf{A. Picksley}: Conceptualization (equal), Data curation (equal), Formal analysis (equal), Methodology (equal), Visualization (equal), Writing - review and editing (equal).
\textbf{C. Benedetti}: Data curation (equal), Formal analysis (equal), Writing - original draft (equal), Writing - review and editing (equal).
\textbf{F. Fillippi}: Data curation (equal), Formal analysis (equal), Writing - review and editing (equal).
\textbf{J. Stackhouse}: Formal analysis (equal), Methodology (equal).
\textbf{L. Fan-Chiang}: Data curation (equal), Methodology (equal).
\textbf{H. E. Tsai}: Formal analysis (equal), Methodology (equal).
\textbf{K. Nakamura}: Project administration (equal), Supervision (equal).
\textbf{C. B. Schroeder}: Project administration (equal), Funding acquisition (lead), Supervision (equal), Writing - review and editing (equal).
\textbf{J. van Tilborg}: Project administration (equal), Funding acquisition (equal), Writing - review and editing (equal).
\textbf{E. Esarey}: Project administration (equal), Funding acquisition (equal).
\textbf{C. G. R. Geddes}: Project administration (equal), Funding acquisition (equal).
\textbf{A. J. Gonsalves}: Conceptualization (lead), Writing - review and editing (equal), Project administration (equal), Funding acquisition (equal), Supervision (lead).

\section*{Data Availability Statement}

The data that support the findings of this study are available from the corresponding author upon reasonable request.

\bibliography{aipsamp}

\end{document}